\documentclass[twocolumn,pra,aps,amsmath,amssymb]{revtex4}
\usepackage{graphicx}
\usepackage{epsfig}

\newcommand{\ket}[1]{\vert #1 \rangle}

\newcommand{\ketbra}[1]{\vert #1 \rangle\langle #1 \vert}
\newcommand{\expect}[1]{\left\langle #1\right\rangle}
\begin{document}
\title{Random multiparty entanglement distillation}
\author{Ben Fortescue\footnote{bfort@physics.utoronto.ca}}
\author{Hoi-Kwong Lo\footnote{hklo@comm.utoronto.ca}}
\affiliation{%
Center for Quantum Information and Quantum Control (CQIQC),
Dept. of Electrical \& Computer Engineering and Dept. of Physics,
University of Toronto, Toronto, Ontario, M5S 3G4, Canada
}%
\date{January 15, 2008}
\begin{abstract}
We describe various results related to the random distillation of multiparty entangled states - that is,
conversion of such states into entangled states shared between fewer parties, where those
parties are not predetermined.  In previous work \cite{FL} we showed that certain output states (namely Einstein-Podolsky-Rosen (EPR) pairs)
could be reliably acquired from a prescribed initial multipartite state (namely the $W$ state
$\ket{W}=\frac{1}{\sqrt{3}}(\ket{100}+\ket{010}+\ket{001})$)
via random distillation that could not be reliably created between predetermined parties.  Here we
provide a more rigorous definition of what constitutes ``advantageous'' random distillation.  We show
that random distillation is always advantageous for $W$-class three-qubit states (but only sometimes
for Greenberger-Horne-Zeilinger (GHZ)-class states).  We show that the general class of multiparty states known as symmetric Dicke states
can be readily converted to many other states in the class via random distillation.  Finally we show
that random distillation is provably not advantageous in the limit of multiple copies of pure states.
\end{abstract}
\maketitle

\section{Introduction}
Entanglement in quantum information theory is often considered as a resource \cite{DHW,mother}  which can
be used by physically separated parties to perform tasks such as quantum teleportation \cite{teleport}
or superdense coding \cite{dcode}, under the restriction of the parties to local operations and classical
communications (LOCC).  Under LOCC the parties can perform local quantum operations on their own
portions of the entangled states and exchange classical information with each other through some classical
communication channels, but not perform joint quantum
operations or (equivalently) exchange quantum information.  It is known that parties cannot increase
their shared entanglement under LOCC, which motivates the view of entanglement as a resource.

Determining what may be accomplished with some particular entangled state under LOCC provides an operational
description of that state, which can in some cases be used as an entanglement measure - for example the
well-known result \cite{BBPS} that the maximum ratio at which maximally-entangled EPR pairs
\begin{equation}
\ket{\Phi}=\frac{1}{\sqrt{2}}(\ket{00}+\ket{11})_{AB}
\end{equation}
can be obtained through LOCC is the entanglement entropy $E^S(\psi_{AB})=S(\rho_A)$, where
\begin{align}
S(\rho)&=-\textrm{tr} \rho\log_2\rho\\
\rho_A&=-\textrm{tr}_B(\ketbra{\psi_{AB}}).
\end{align}

In general, the properties of multiparty entangled states (those shared between more than two parties)
are much less-well understood than those of two-party states.  For example, there is no single well-defined
maximally entangled state in the multiparty case.  It appears that multiparty
states can divided into distinct classes \cite{dur,VDMV}, and even in the three-party case
it is not known whether or not all entangled states can be reversibly obtained through LOCC from a finite
selection of other states - the ``minimal reversible entanglement generating set'' (MREGS) \cite{BPRSV}.

A topic of interest in the description of multiparty entangled states 
is the conversion of these states into, generally, states shared between
fewer parties, and specifically two-party states.  Since there are many
results on the operational properties of two-party states, considering such
a conversion provides useful information in the multiparty case.

Several results e.g. \cite{ea, LVE, SVW, horo, Gour, GS} exist regarding the conversion
of multiparty to two-party entangled states shared between predetermined parties.  In \cite{FL}
we demonstrated that some two-party entangled states which could not be reliably obtained (i.e. probability $<1$) between
predetermined parties could be reliably obtained (probability $\to 1$ in the limit of many ``rounds'' of distillation) between parties which were randomly
determined in the course of a LOCC-protocol - a process we refer to as ``random distillation''.
Specifically
we showed that one can reliably distill one EPR pair from a single $W$ between random parties, versus
doing so with a probability at most 2/3 between predetermined parties.  The random distillation rate exceeds even the
asymptotic rate of $H_2(1/3)\approx 0.92$ EPRs per $W$ between predetermined parties in the many-copy limit,
where $H_2$ is the binary entropy function
\begin{equation}
H_2(x)=-x\log_2(x)-(1-x)\log_2(1-x).
\end{equation}

In this paper, we address a number of questions in random distillation.  Firstly. our criterion
in \cite{FL} for what constituted ``advantageous'' random distillation was somewhat
problematic, in particular when considering multiple copies of states.  Here we provide
a new criterion for advantageous random distillation applicable to any pure-state case,
including that of collective operations on multiple copies of a state.  We also ask whether
random distillation gives an advantage in the many-copy limit, and demonstrate that it does not.

Secondly, in our previous paper we considered only a small number of specific states.
Here we consider the random distillation properties of general classes of states - specifically, 
distilling the general classes of three-qubit pure-state entanglement,
the GHZ and $W$ classes.  We show that all $W$-class states can undergo advantageous random distillation,
but the GHZ class contains examples both of states which can and cannot.

Finally, we previously considered primarily distillation to two-party EPR pairs.  Here
we consider a class of final states shared between larger numbers of parties.
For the multiparty entangled states known as symmetric Dicke states, we briefly demonstrate
a class of output states which may be reliably obtained through LOCC only by random distillation.

\section{Definitions}
For conversion of an $M$-party pure state $\ket{\psi}$ to EPR pairs $\ket{\Phi}$
through LOCC
\begin{equation}
\ket{\psi}^{\otimes N}_{A_1\ldots A_M}\underbrace{\longrightarrow}_{LOCC} 
\bigotimes_{ij}\ket{\Phi}_{A_iA_j}^{\otimes N_{A_iA_j}}\label{eq-distill}.
\end{equation}
we define
\begin{align}
E^\infty_{a_{IJ}}(\psi)&\equiv\sup_{N\to\infty}\frac{N_{A_IA_J}}{N}\\
E^\infty_{s}(\psi)&\equiv\max_{ij}\sup_{N\to\infty}\frac{N_{A_IA_J}}{N}\\
E^\infty_t(\psi)&\equiv\sup_{N\to\infty}\frac{\sum_{ij}N_{A_iA_j}}{N}.
\end{align}
That is, $E^\infty_{a_{IJ}}$ represents the maximum rate of EPR distillation between parties $I$ and $J$ (with the help of all other
parties),
$E^\infty_{s}$ represents the highest distillation rate of EPR pairs between any given pair of parties
and $E^\infty_t$ the highest total EPR distillation rate, irrespective of which parties share them.

In this asymptotic case (though not generally) $E^\infty_{a_{IJ}}$ is equal to the entanglement of assistance
\cite{ea}, with \cite{horo} showing that
\begin{equation}
E^\infty_{a_{IJ}}=\min_T \{S(\rho_{A_I T}), S(\rho_{A_J\overline{T}})\}\label{eq-horo}
\end{equation}
where the minimization is over the division of parties other than $A_I$ and $A_J$ into two groups $T$ and $\overline{T}$ (i.e. over
bipartite ``cuts'' separating all parties into two groups, one containing $A_I$ and one containing $A_J$) and
\begin{equation}
\rho_{A_IT}=\textrm{tr}_{A_{j\notin\{I,T \}}}(\ketbra{\psi}),
\end{equation}
the reduced state of $\ketbra{\psi}$, traced over all $A_{j\notin \{I,T\}}$.

For the single-copy analogues of these quantities, for the distillation
\begin{equation}
\ket{\psi}_{A_1\ldots A_M}\underbrace{\longrightarrow}_{LOCC} 
\bigotimes_{ij}\ket{\Phi}_{A_iA_j}^{\otimes N_{A_iA_j}}\label{eq-distill},
\end{equation}
we define
\begin{align}
E_{a_{IJ}}(\psi)&\equiv\sup\expect{N_{A_IA_J}}\\
E_{s}(\psi)&\equiv\max_{ij}\left(\sup\expect{N_{A_IA_J}}\right)\\
E_{t}(\psi)&\equiv\sup\expect{\sum_{ij}N_{A_iA_j}}
\end{align}

We noted in \cite{FL} that for the three-party $W$ state
\begin{equation}
\ket{W}=\frac{1}{\sqrt{3}}(\ket{100}+\ket{010}+\ket{001})_{ABC}
\end{equation}
it is possible to obtain an EPR through LOCC between random parties but not specified parties.  Hence even though
(from \cite{horo}) $E^\infty_{s}(W)=H_2(1/3)\approx 0.92$,
we find $E^\infty_t(W)\ge 1$.  However \cite{FL} further noted that the condition $E_t> E_{s}$ could
also be trivially satisfied, for example by the state $\ket{\Phi}_{AB}\otimes\ket{\Phi}_{BC}$, for which $E_t=2>E_{s}=1$.

We would therefore like to find a condition that more generally captures when true ``random distillation'' is advantageous
 - that is, one obtains a greater entanglement yield due to the nondeterministic nature (in terms of which parties receive the final state) of the distillation , rather than
there simply being somewhat independent entanglements between different pairs of parties.  We would further like
to define such a condition in terms of general pure-state bipartite entanglement measures, rather than solely
in terms of the distillable EPR pairs.

We thus consider the LOCC-conversion (via a protocol $P$) of an initial pure state $\psi$ to final pure multipartite states
$\psi_f$ with probabilities $p_f$.
\begin{equation}
\psi\underbrace{\xrightarrow{P}}_{LOCC}\{\psi_f, p_f\}\label{eq-locc}
\end{equation}
and the LOCC conversion (via a protocol $Q$) of multiparty states $\psi_f$ to pure two-party states $\psi_{g_{IJ}}$ with probabilities
$p_g$
\begin{equation}
\psi_f\underbrace{\xrightarrow{Q}}_{LOCC}\{\psi_{g_{IJ}}\otimes \rho_g, p_g\}
\end{equation}
(note that in the above, $I$ and $J$ are not necessarily the same for every $g$).

We define, for some bipartite pure-state entanglement measure $E$
\begin{align}
A_{IJ}(\psi_f)&\equiv\sup_{P,Q}\sum_g p_gE(\psi_{g_{IJ}})\\
E_{sp}(\psi_f)&\equiv\max_{IJ} A_{IJ}(\psi_f)\\
E_{rnd}(\psi)&\equiv\sup_{P,Q}\sum_f p_fE_{sp}(\psi_f)\label{eq-ernd}\\
E^\infty_{rnd}(\psi)&\equiv\frac{E_{rnd}(\psi^{\otimes N})}{N},\quad N\to\infty\label{eq-erndi}\\
E^\infty_{sp}(\psi)&\equiv\frac{E_{sp}(\psi^{\otimes N})}{N},\quad N\to\infty\label{eq-espi}
\end{align}
where the supremums in the above expressions are over all possible LOCC protocols $P$ and $Q$.

Hence $E_{rnd}$ represents the supremum of the expected entanglement (as measured by $E$) obtained by whichever
pair of parties has the highest entanglement once the protocol has been performed, while $E_{sp}$
is the corresponding quantity for parties chosen before performing the protocol.  Thus $E_{rnd} \ge E_{sp}$ in
general, and, if $E_{rnd}(\psi)>E_{sp}(\psi)$, this represents genuine advantageous random distillation
as discussed above.

While as mentioned any bipartite pure-state measure $E$ may in principle be used, for the remainder of this paper
and our results (with the exception of Section \ref{sec-dicke}) we shall adopt as our measure the entanglement entropy $E^S$, i.e.
the Von Neumann entropy $S$ of the reduced state as noted above.  We thus define
\begin{equation}
E(\psi_{AB})\equiv E^S(\psi_{AB}).
\end{equation}

\section{W and GHZ-class states}

In \cite{FL} we demonstrated advantageous random distillation for the three-party $W$ and similar states, and that
random distillation was not advantageous for certain GHZ-like states.  However no general
result was obtained for the general GHZ and W classes noted in \cite{dur}, into one of which any three-qubit
pure state with genuine tripartite entanglement may be classed.  Here we find\\

\noindent \textbf{Theorem 1:}\\
For any W-class pure entangled three-qubit state $\psi_W$
\begin{equation}
E_{rnd}(\psi_W)>E_s(\psi_W)
\end{equation}
\noindent \textbf{Proof:}\\
We make use of the following simple lemma\\

\noindent{\bf Lemma 1:}
For a general normalised two-qubit pure state
\begin{equation}
\psi=(c_{00}\ket{00}+c_{01}\ket{01}+c_{10}\ket{10}+c_{11}\ket{11})_{AB}
\end{equation}
the entanglement measure $S(\rho_A)$ increases monotonically
with the concurrence \cite{wooters}
\begin{equation}
q(\psi)=2|c_{01}c_{10}-c_{00}c_{11}|\label{eq-qdef}
\end{equation}
and $S(\rho_A)$ is a convex function of $q(\psi)$ in the range $0\le q \le 1$
, corresponding to $0\le S \le 1$.\\

\noindent{\bf Proof:}
Explicit calculation shows
\begin{equation}
S(\rho_A)=f(q)=H_2\left(\frac{1-\sqrt{1-q(\psi)^2}}{2}\right)\label{eq-sq}
\end{equation}
and that
\begin{equation}
\frac{d^2 f}{dq^2}\ge 0,\quad 0\le S\le 1\quad \Box
\end{equation}

We define $q_{sp}$, $q_{rnd}$ etc. as analogous quantities to $E_{sp}, E_{rnd}$ etc., with $q$ as the entanglement measure.
The quantity $q$ is a useful measure in this case since it is second-order in the state's coefficients. Thus, for repeated rounds of unitaries and measurements,
probabilities and normalisation factors cancel out when calculating $\expect{q}$, as shown below.  Since the $E_x$ (i.e. $E_{rnd}$, $E_{sp}$ etc.)
are expectation values for $S$,
it follows from the convexity result that
\begin{equation}
E_x(\psi)\ge f(q_x(\psi))\label{eq-convex}.
\end{equation}
Note then that by this definition $q_x(\psi)\ne f^{-1}(E_x(\psi))$, in general.

\subsection{The $W$ protocol}\label{sec-rd}
We first consider the protocol of \cite{FL} (which we will refer to as the $W$ protocol)
for obtaining an EPR pair from a $W$ state.
This consists of all three parties repeatedly applying the unitary
\begin{equation}
\ket{0}\longrightarrow \sqrt{1-\epsilon^2}\ket{0}+\epsilon\ket{2},\quad \ket{1}\longrightarrow\ket{1}\label{eq-uni}
\end{equation}
followed by all performing the projection
\begin{equation}
F=\ketbra{0}+\ketbra{1},\quad G=\ketbra{2}.\label{eq-project}
\end{equation}
If all three parties get outcome $F$, the protocol is repeated. If exactly one party gets outcome $G$, the
other two parties have an EPR pair, the expectation value of their eventual entanglement tending to unity in the limit
of many repetitions and small $\epsilon$.  (The probability of the protocol aborting due to
failure, where two or more parties get $G$, is negligible in this limit).

(We also show in \cite{FL} that random distillation is advantageous for a finite number of rounds, with a protocol
for which the probability of obtaining a randomly-shared EPR pair from a $W$ within $R$ rounds is $\frac{R}{R+1}$.  This exceeds
the single-copy limit (for predetermined parties) of 2/3 for $R\ge 3$ and the asymptotic limit of 0.92 for
$R\ge 12$.)

Note that the $W$ state enjoys a special property that makes our previous analysis of random distillation
of an EPR from a $W$ state simple - a failed round (that is, where all parties obtain outcome $F$) returns
the state to a $W$.  Therefore in the limit of many rounds and small $\epsilon$ (where success and failure
of this kind are the only outcomes with non-negligible probability) the protocol is ``reset'' after
each failure and every round can be analysed in the same way.  In contrast, this is {\it not} the case
for a general three-qubit pure state.  Indeed, whenever a round of random distillation fails a general
three-qubit state becomes a new state.  For this reason, the analysis of multi-round random distillation
for the general three-qubit state is not entirely trivial.  In the following, we will use the properties
of the concurrence discussed above to perform such an analysis.  Before doing so, let us first demonstrate the evolution
of a general three-qubit state under the $W$ protocol.

Consider then applying this protocol to a general three-qubit pure state shared between Alice, Bob and Charlie:
\begin{multline}
\ket{\psi_1}_{ABC}=\ket{0}_A\Big(k_{{00}_0}\ket{00}+k_{{01}_0}\ket{01}\\
+k_{{10}_0}\ket{10}+k_{{11}_0}\ket{11}\Big)_{BC}
+\ket{1}_{A}(\ldots)\label{eq-state}
\end{multline}
where the $(\ldots)$ represent some additional terms whose amplitudes we are not concerned with.
We define $K_{{00}_0}\equiv|k_{{00}_0}|^2$ etc.

After every party has performed the unitary (\ref{eq-uni}) the state becomes
\begin{multline}
\ket{\psi_1}_{ABC}=(1-\epsilon^2)^{\frac{1}{2}}\ket{0}_A\Big((1-\epsilon^2)k_{{00}_0}\ket{00}\\
+(1-\epsilon^2)^{\frac{1}{2}}[k_{{01}_0}\ket{01}+k_{{10}_0}\ket{10}]+k_{{11}_0}\ket{11}\Big)_{BC}\\
+\epsilon\ket{2}_A\Big((1-\epsilon^2)k_{{00}_0}\ket{00}+(1-\epsilon^2)^{\frac{1}{2}}[k_{{01}_0}\ket{01}+k_{{10}_0}\ket{10}]\\
+k_{{11}_0}\ket{11}\Big)_{BC}
+(\ldots).
\end{multline}
If all the parties then perform the projection (\ref{eq-project}) and all get outcome $F$ the resultant state
will differ from the initial state.  Likewise if these unitaries and projections repeat until Alice, say, eventually gets outcome $G$ the state then shared by
Bob and Charlie will depend on the number of rounds performed up to that point.

In general after $R$ rounds of unitaries and projections in which all parties get $F$, the shared state will be
\begin{multline}
\ket{\psi_R}_{ABC}=\ket{0}_A\Big(k_{{00}_R}\ket{00}+k_{{01}_R}\ket{01}\\
+k_{{10}_R}\ket{10}+k_{{11}_R}\ket{11}\Big)_{BC}
+\ket{1}_{A}(\ldots)_{BC}
\end{multline}
where
\begin{align}
k_{{00}_R}&=\frac{(1-\epsilon^2)^\frac{3R}{2} k_{{00}_0}} {\sqrt{P_{F_R}\ldots P_{F_1}}}\\
k_{{01}_R}&=\frac{(1-\epsilon^2)^{R}    k_{{01}_0}} {\sqrt{P_{F_R}\ldots P_{F_1}}}\\
k_{{10}_R}&=\frac{(1-\epsilon^2)^{R}    k_{{10}_0}} {\sqrt{P_{F_R}\ldots P_{F_1}}}\\
k_{{11}_R}&=\frac{(1-\epsilon^2)^\frac{R}{2}  k_{{11}_0}} {\sqrt{P_{F_R}\ldots P_{F_1}}}
\end{align}
and $P_{F_N}$ is the probability of all parties getting $F$ in the $N$th round of the protocol
after having done so in all previous rounds i.e.
\begin{multline}
P_{F_N}=(1-\epsilon^2)\Big((1-\epsilon^2)^2 K_{{00}_{N-1}}\\
+(1-\epsilon^2)[K_{{01}_{N-1}}+K_{{10}_{N-1}}]+K_{{11}_{N-1}}\Big)
\end{multline}
If the parties perform one further round of unitaries, the state will be
\begin{multline}
\ket{\psi_{R+1}}_{ABC}=(1-\epsilon^2)^{\frac{1}{2}}\ket{0}_A\Big((1-\epsilon^2)k_{{00}_R}\ket{00}\\
+(1-\epsilon^2)^{\frac{1}{2}}[k_{{01}_R}\ket{01}
+k_{{10}_R}\ket{10}]+k_{{11}_R}\ket{11}\Big)_{BC}\\
\shoveleft{+\epsilon\ket{2}_A\Big((1-\epsilon^2)k_{{00}_R}\ket{00}}\\
+(1-\epsilon^2)^{\frac{1}{2}}[k_{{01}_R}\ket{01}
+k_{{10}_R}\ket{10}]
+k_{{11}_R}\ket{11}\Big)_{BC}\\
+(\ldots)
\end{multline}
If the parties then project and Alice alone gets outcome $G$, with probability
\begin{equation}
P_{G_{R+1}}=\epsilon^2\Big((1-\epsilon^2)^2 K_{{00}_R}+(1-\epsilon^2)[K_{{01}_R}+K_{{10}_R}]+K_{{11}_R}\Big)
\end{equation}
the resultant state will be
\begin{multline}
\frac{1}{\sqrt{P_{G_{R+1}}}}\epsilon\ket{2}_A\Big((1-\epsilon^2)k_{{00}_R}\ket{00}\\
+(1-\epsilon^2)^{\frac{1}{2}}[k_{{01}_R}\ket{01}+k_{{10}_R}\ket{10}]+k_{{11}_R}\ket{11}\Big)_{BC}
\end{multline}
and Bob and Charlie will share a state with entanglement (measured by the concurrence $q$ (\ref{eq-qdef}))
\begin{align}
q_{R+1}^{BC}=&\frac{1}{P_{G_{R+1}}}\epsilon^2(1-\epsilon^2)\times 2|k_{{01}_R}k_{{10}_R}-k_{{00}_R}k_{{11}_R}|\\
=&\frac{2}{P_{G_{R+1}}P_{F_R}\ldots P_{F_1}}\epsilon^2(1-\epsilon^2)^{2R+1}\nonumber\\
&\times|k_{{01}_0}k_{{10}_0}-k_{{00}_0}k_{{11}_0}|
\end{align}

Thus if we consider applying the $W$ protocol to an arbitrary three-qubit state we have that for the final
expected concurrence $\expect{q_f^{BC}}$ (\ref{eq-qdef}):

\begin{align}
\expect{q_f^{BC}}&\ge \lim_{\epsilon\to 0} \sum_{R=0}^\infty q_{R+1}^{BC} P_{G_{R+1}}\prod_{N=1}^R P_{F_N}\\
&= 2|k_{{01}_0}k_{{10}_0}-k_{{00}_0}k_{{11}_0}|\times\lim_{\epsilon\to 0}\sum_{R=0}^\infty \epsilon^2(1-\epsilon^2)^{2R+1}\\
&=2|k_{{01}_0}k_{{10}_0}-k_{{00}_0}k_{{11}_0}|\times\lim_{\epsilon\to 0}\frac{\epsilon^2(1-\epsilon^2)}{1-(1-\epsilon^2)^2}\\
&=|k_{{01}_0}k_{{10}_0}-k_{{00}_0}k_{{11}_0}|.
\end{align}

The above bound concerns only Bob and Charlie's entanglement as a result of Alice eventually getting outcome $G$ (and the
others $F$).  
However other possible outcomes are where instead Bob or Charlie gets $G$ resulting in zero Bob-Charlie entanglement,
but some entanglement between Alice-Bob or Alice-Charlie.  How much entanglement depends on the form
of the original state, but since the $W$ protocol is symmetric (i.e. invariant with respect to permutation of parties), we see
that in the special case of a symmetric state $\psi^{symm}_{ABC}$, the expected entanglement due to such outcomes
must also be $|k_{{01}_0}k_{{10}_0}-k_{{00}_0}k_{{11}_0}|=|k_{{01}_0}^2-k_{{00}_0}k_{{11}_0}|$ (since
$k_{{01}_0}=k_{{10}_0}$ for symmetric $\psi_{ABC}$), for each of Alice-Bob and Alice-Charlie.

Thus, considering only these outcomes where two parties share some entanglement
and are unentangled with the third party, it follows that

\begin{equation}
E_{rnd}(\psi^{symm}_{ABC}) \ge 3|k_{{01}_0}^2-k_{{00}_0}k_{{11}_0}|
\end{equation}

\cite{dur} showed that a general $W$-class state could be expressed as
\begin{equation}
(\alpha\ket{100}+\beta\ket{010}+\gamma\ket{001}+\delta\ket{000})_{ABC}\label{eq-Wclass}
\end{equation}
where $\{\alpha,\beta,\gamma,\delta\}\in \mathbb{R}$ and $\alpha,\beta,\gamma>0$, $\delta\ge 0$.  We will without loss of
generality take $\gamma\ge\beta\ge\alpha$.

We find for the state (\ref{eq-Wclass}) that
\begin{align}
S(\rho_A)=H_2(\lambda)\textrm{, where}\\
\lambda^2-\lambda+\alpha^2(\beta^2+\gamma^2)=0
\end{align}
Using (\ref{eq-sq}), we find the corresponding concurrences
\begin{align}
q(\rho_A)&=2\alpha\sqrt{\beta^2+\gamma^2}\\
q(\rho_B)&=2\beta\sqrt{\alpha^2+\gamma^2}\label{eq-rb}\\
q(\rho_C)&=2\gamma\sqrt{\alpha^2+\beta^2}
\end{align}
It is straightforward to see that $q(\rho_C)\ge q(\rho_B)\ge q(\rho_A)$ and thus
\begin{equation}
E^\infty_{sp}(\psi_W)=S(\rho_B).
\end{equation}
\subsection{A random distillation for W-class states}
We will see that a higher entanglement than the above may be obtained for a $W$-class state by first symmetrising
it and then performing random distillation via the $W$ protocol.
Starting with the state (\ref{eq-Wclass}) Alice applies the unitary
\begin{equation}
\ket{0}\longrightarrow \frac{\alpha}{\gamma}\ket{0}+\sqrt{1-\left(\frac{\alpha}{\gamma}\right)^2}\ket{2},\quad \ket{1}\longrightarrow\ket{1}
\end{equation}
producing the state
\begin{align}
\left(\alpha\ket{100}+\frac{\beta\alpha}{\gamma}\ket{010}+\alpha\ket{001}+\frac{\delta\alpha}{\gamma}\ket{000}\right)_{ABC}\nonumber\\
+\sqrt{1-\left(\frac{\alpha}{\gamma}\right)^2}\ket{2}_A\left(\beta\ket{10}+\gamma\ket{01}+\delta\ket{00}\right)_{BC}
\end{align}
Alice then projects using (\ref{eq-project}).  If she receives outcome $G$ (with probability $1-P_{AF}$) the protocol terminates, otherwise
Bob then applies the unitary
\begin{equation}
\ket{0}\longrightarrow \frac{\beta}{\gamma}\ket{0}+\sqrt{1-\left(\frac{\beta}{\gamma}\right)^2}\ket{2},\quad \ket{1}\longrightarrow\ket{1}
\end{equation}
producing the state
\begin{align}
\frac{1}{\sqrt{P_{AF}}}\Bigg[\left(\frac{\alpha\beta}{\gamma}(\ket{100}+\ket{010}+\ket{001})+\frac{\delta\alpha\beta}{\gamma^2}\ket{000}\right)_{ABC}\\
+\sqrt{1-\left(\frac{\beta}{\gamma}\right)^2}\ket{2}_B\left(\alpha\ket{10}+\alpha\ket{01}+\frac{\delta\alpha}{\gamma}\ket{00}\right)_{AC}\Bigg]
\end{align}
Bob likewise then projects using (\ref{eq-project}), the protocol terminating if he gets outcome $G$.  If he gets outcome $F$
(conditional probability $P_{BF}$),
the state obtained is
\begin{equation}
\frac{1}{\sqrt{P_{AF}P_{BF}}}\frac{\alpha\beta}{\gamma}\left(\ket{100}+\ket{010}+\ket{001}+\frac{\delta}{\gamma}\ket{000}\right)_{ABC}
\end{equation}
which is a symmetric state on which the three parties perform the $W$ protocol.

Thus for the overall protocol
\begin{align}
q_{rnd}(\psi_W)&\ge(1-P_{AF})\times 2\frac{\left(1-\left(\frac{\alpha}{\gamma}\right)^2\right)\beta\gamma}{1-P_{AF}}\nonumber\\
&+P_{AF}(1-P_{BF})\times 2\frac{\left(1-\left(\frac{\alpha}{\beta}\right)^2\right)\alpha^2}{P_{AF}(1-P_{BF})}\nonumber\\
&+P_{AF}P_{BF}\times 3\frac{\left(\frac{\alpha\beta}{\gamma}\right)^2}{P_{AF}P_{BF}}\nonumber\\
&=2\left(1-\frac{\alpha^2}{\gamma^2}\right)\beta\gamma+2\alpha^2+\frac{\alpha^2\beta^2}{\gamma^2}\label{eq-qrnd}
\end{align}

We use the Lemma\\

\noindent{\bf Lemma 2:}
\begin{align}
q_{rnd}(\psi_W)=&2\left(1-\frac{\alpha^2}{\gamma^2}\right)\beta\gamma+2\alpha^2+\frac{\alpha^2\beta^2}{\gamma^2}\nonumber\\
>&q(\rho_B)=2\beta\sqrt{\alpha^2+\gamma^2}
\end{align}

\noindent{\bf Proof:}
See Appendix A.\\

Hence from (\ref{eq-convex})
\begin{equation}
E_{rnd}(\psi_W)\ge f(q_{rnd}(\psi_W))>f(q(\rho_B))=E_{sp}(\psi_W).\quad \Box
\end{equation}

\subsection{GHZ-class states}
As noted in \cite{FL}, the above inequality ($E_{rnd}(\psi)>E_{sp}(\psi)$) is not generally true for GHZ class states, with the GHZ state itself
, and more generally states of the form $\alpha\ket{000}+\beta\ket{111}$ (for which $E_{sp}=E_{rnd}$) providing a counterexample.  One might wonder
whether random distillation gives no advantage for any state in the GHZ class.  Here, we answer this question in the negative.  More specifically,
we find an explicit example of a GHZ class state for which random distillation gives an advantage over distillation to predetermined parties.

Our example state is
\begin{equation}
\ket{\psi_G}=\alpha(\ket{100}+\ket{010}+\ket{001})+\epsilon\ket{111},\quad \epsilon=\sqrt{1-3\alpha^2}.
\end{equation}
for $0<\{\alpha,\beta,\gamma,\delta,\epsilon\}\in\mathbb{R}$.
The three-tangle $\tau_{ABC}$ \cite{CKW} for this state is equal to $16\epsilon\alpha^3$, and being non-zero
the state is thus \cite{dur} GHZ-class.

By symmetry of $\psi_G$, we have $E_{sp}(\psi_G)=H_2(\alpha^2+\epsilon^2)$, and
\begin{equation}
f^{-1}(E_{sp}(\psi_G))=\sqrt{8\alpha^2(1-2\alpha^2)}
\end{equation}
From its symmetry and the analysis of section \ref{sec-rd}, $\psi_G$ can be randomly distilled
to obtain
\begin{equation}
q_{rnd}=3\alpha^2.
\end{equation}
It follows that  $q_{rnd}(\psi_G)>f^{-1}(E_{sp}(\psi_G))$ and hence $E_{rnd}(\psi_G)>E_{sp}(\psi_G)$ for $\alpha^2>8/25$.
I.e. there exist GHZ class states for which random distillation is
advantageous and (as shown in \cite{FL}) others for which it is not.

\section{Symmetric Dicke states}\label{sec-dicke}
While we do not have a general treatment of random distillation applied to
pure states shared between $> 3$ parties, it is clear that there are such states
from which final states shared between fewer parties can be reliably obtained iff those
parties are not predetermined.  In \cite{FL} we gave the example of the $M$-party $W$ state,
(a symmetric superposition of the $M$-qubit states with a single excited qubit)
\begin{equation}
\ket{W_M}=\frac{1}{\sqrt{M}}(\ket{0\ldots 01}+\ket{0\ldots 010}+(\textrm{permutations}))
\end{equation}
to which applying the $W$ protocol produces a randomly-shared $W_{M-1}$ state.
Considering a bipartite split of the initial and final states between
one of the parties $P$ who shares the final state and all other parties, we see
that
\begin{equation}
S(\sigma_{Pf})=H_2\left(\frac{1}{M-1}\right)>S(\sigma_{Pi})=H_2\left(\frac{1}{M}\right)
\end{equation}
where $i$ and $f$ denote initial and final states.
Thus such a distillation cannot be reliably performed for predetermined final parties.

We can also consider a more general class of states whose entanglement properties are of some interest \cite{KSTSW,toth,symm}
- the $M$-party symmetric Dicke states \cite{Dicke, MW}.
These are of the form
\begin{equation}
\ket{\psi(M,N)}=\frac{1}{\sqrt{{^M}C_N}}\sum \ket{\textrm{$N$ 1s, $(M-N)$ 0s}}
\end{equation}
where ${^M}C_N$ are the binomial coefficients
\begin{equation}
{^M}C_N\equiv \frac{M!}{N!(M-N)!}
\end{equation}
and the sum is over all permutations of the individual qubits.  E.g.
\begin{align}
\ket{\psi(4,2)}=\frac{1}{\sqrt{6}}(&\ket{0011}+\ket{0110}+\ket{1100}\nonumber\\
&+\ket{1001}+\ket{0101}+\ket{1010}).
\end{align}
Considering the Von Neumann entropy of a party $P$ we have
\begin{equation}
S(\sigma^{M,N}_P)=H_2 \left(\frac{1}{{^M}C_N}\right)
\end{equation}
and hence any LOCC distillation $\psi(M,N)\longrightarrow \psi(M',N')$ cannot be reliably
performed for predetermined final parties if ${^{M'}}C_{N'}<{^M}C_N$.

However, we see that if we apply the $W$ protocol to a state $\psi(M,N)$
we can reliably obtain either a randomly-shared $\psi(M-1,N)$ (applying the usual protocol)
or $\psi(M-1,N-1)$ (applying the $W$ protocol but with $\ket{0}$ and $\ket{1}$ states reversed).
Essentially the parties can reliably "drop" either a $\ket{0}$ or $\ket{1}$ from the
terms of the state to produce a state randomly shared between one fewer party.

Given that the parties can also (by all applying a bit-flip operation) always reliably
convert $\psi(M,N)\longrightarrow\psi(M,M-N)$, we find that the parties can reliably perform
\begin{align}
\ket{\psi(M,N)}&\longrightarrow\ket{\psi(M',N')},\textrm{ or }\ket{\psi(M',M'-N')}\textrm{ if }\\
M'&\le M\nonumber\\
N'&\ge(M'-M)+N\nonumber
\end{align}
many of which output states could not be achieved for predetermined final parties.

\section{Random distillation in the many-party limit}
In our previous paper \cite{FL}, we show that random distillation is useful for the case of a single copy of the $W$ state.
One might wonder whether random distillation remains advantageous in the limit of many copies of a general pure state
(including $W$ states).  Here we show that (according to our current definition) the answer is no.

In \cite{FL} we showed that one could randomly distill one EPR pair from a single $W$ state compared to 0.92 EPRs per $W$ between predetermined parties in the many-copy limit.
Trivially, it follows that for multiple copies of the $W$ state we can obtain advantageous random distillation in the context of $E_t>E_{sp}$ - that is,
many copies of the $W$ state can produce more EPR pairs {\it in total} (summing up those
between all pairs of parties) than can be obtained between predetermined parties.

However, this does not tell us whether random distillation remains useful for many copies of a pure state in our redefined sense of
$E_{rnd}>E_{sp}$ - obtaining more entanglement between only two parties when the two are not predetermined.

In what follows, we will discuss the case of two copies of $W$ states
and note that we find an advantage for random distillation for this case.
More concretely, we can easily devise a simple two-copy analogue to the $W$ protocol, in which three parties sharing two $W$ states
each repeatedly perform the two-qubit unitary
\begin{equation}
\ket{00}\longrightarrow\sqrt{1-\epsilon^2}\ket{00}+\epsilon\ket{2}
\end{equation}
(with all other states ($\ket{01},\ket{10},\ket{11}$) mapping to themselves) combined
with a projection into either a $\ket{2}$ state or the $SU(2)\otimes SU(2)$ subspace.
As with the general three-qubit state, in this case repeated rounds change the overall
state.  We find, by considering a four-qubit measure analogous to $q$, that
\begin{align}
E_{rnd}(W^{\otimes 2})&\ge -2[\zeta\log_2\zeta-(0.5-\zeta)\log_2(0.5-\zeta)]\nonumber\\
&\approx 1.843\textrm{, where}\\
\zeta&=\frac{1-\sqrt{1-\left(\frac{8}{9}\right)^2}}{4}
\end{align}
Hence
\begin{equation}
E_{rnd}(W^{\otimes 2})>E^\infty_{sp}(W^{\otimes 2})=2H_2\left(\frac{1}{3}\right)\approx 1.837.
\end{equation}
Hence there is an advantage to random distillation of $W^{\otimes 2}$, but the proven advantage is very marginal.
We see that this extending this protocol in a na{\" i}ve manner to more copies (i.e
performing a unitary $\ket{0}^{\otimes N}\to \sqrt{1-\epsilon^2}\ket{0}^{\otimes N}+\epsilon\ket{2}$ etc.)
will not sustain the advantage, since for $N$ copies the probability of success will fall roughly as $O(\frac{1}{3^N})$,
while not predetermining the parties will at most triple the expected entanglement.

Confirming this idea more generally, we find in the limit of large $N$:\\

\noindent \textbf{Theorem 2:}\\
\begin{equation}
E_{rnd}(\psi^{\otimes N})\longrightarrow E_{sp}(\psi^{\otimes N}),\quad N\to\infty.
\end{equation}
In other words, as defined in (\ref{eq-erndi}) and (\ref{eq-espi}),
\begin{equation}
E^\infty_{rnd}(\psi)=E^\infty_{sp}(\psi).
\end{equation}

\noindent \textbf{Proof:}\\
This is shown by the result of \cite{LP}, that for a LOCC protocol distilling EPR pairs from $N$ copies of a two-party pure state $\sigma_{AB}$,
\begin{equation}
\ket{\sigma}_{AB}^{\otimes N}\underbrace{\longrightarrow}_{LOCC} \ket{\Phi}_{AB}^{N'}
\end{equation}
the probability of getting $N'>NS(\rho_A)$ tends to 0 as $N\to\infty$.  Specifically the probability shrinks
as $\exp(O(-N))$.  Note that this is stronger than
the well-known result that optimally $\expect{N'}=NS(\rho_A)$, since it disallows improving on the optimum expected yield even
some of the time.

Consider a process (\ref{eq-locc}), where $\psi=\phi_{A_1\ldots A_m}^{\otimes N}$, for some pure state $\phi$.  The optimum
distillation to specified parties will be to some pair of parties $A_I, A_J$.
where (from (\ref{eq-horo})) the distillation rate is $S_{\phi}(A_IT^{\phi}_{IJ})$ where $S_\phi$ denotes Von Neumann entropy of the bracketed parties' reduced state of $\phi$, $T_{IJ}$ in general represents some group of parties not containing $A_I$ or $A_J$ and
$T^{\phi}_{ij}$ is the group that minimises $S_\phi(A_iT_{ij})$, i.e. for any fixed but arbitrary pair of parties $A_i$, $A_j$.
\begin{equation}
S_{\phi}(A_iT^{\phi}_{ij})\le S_{\phi}(A_iT_{ij})\quad\forall\quad T_{ij}.\label{eq-grpmin}
\end{equation}
Thus, as $N\to\infty$
\begin{equation}
E_{sp}(\psi)\longrightarrow NS_{\phi}(A_IT^{\phi}_{IJ})
\end{equation}
and 
\begin{equation}
S_{\phi}(A_IT^{\phi}_{IJ})\ge S_{\phi}(A_iT^\phi_{ij}) \quad\forall\quad ij\label{eq-iparty}
\end{equation}

For $E_{rnd}(\psi)>E_{sp}(\psi)$, by the definition in (\ref{eq-ernd}) we require at least one possible output
state $\psi_f$ to have $E_{sp}(\psi_f)>E_{sp}(\psi)$.  Let us consider one such $\psi_f$, denoted by $\psi_f'$, and occurring
with some fixed probability $p_f'$.  Suppose optimal distillation of $\psi_f'$ (to specified parties) is to parties $A_X$ and $A_Y$ with the corresponding
bipartite cut being between $A_X T_{XY}^f$ on one side  (using, here and below, $f$ to denote quantities defined for reduced states of $\psi_f'$, analogously to $\phi$ above)
and its complementary set on the other side.
Similar to Eqs. (\ref{eq-grpmin}) and (\ref{eq-iparty}), we have for each fixed but arbitrary
pair $i, j$, $S_f (A_i T_{ij}^f ) \leq S_f (A_i T_{ij} ) $ for all $T_{ij}$
and $S_f (A_X T_{XY}^f ) \geq S_f (A_i T_{ij}^f) $ for all $i, j$.  Then, in the many-copy limit
\begin{equation}
E_{sp}(\psi_f')=S_f(A_X T^f_{XY})>E_{sp}(\psi)=NS_{\phi}(A_IT^{\phi}_{IJ})\label{eq-sineq}
\end{equation}

However, from (\ref{eq-grpmin}), (\ref{eq-iparty}) and (\ref{eq-sineq}) we have that
\begin{multline}
S_f(A_XT^\phi_{XY})\ge S_f(A_XT^f_{XY})\\
>NS_{\phi}(A_IT^{\phi}_{IJ})\ge NS_{\phi}(A_XT^{\phi}_{XY})
\end{multline}

Consider now a bipartite division
of $\psi$ between the group $A_XT^\phi_{XY}$ acting as a single party (i.e. we allow joint quantum operations within this group) denoted by $A$
and the group of all other parties acting as a single party $B$.  $A$ and $B$ perform the above LOCC protocol independently
on $M$ copies of $\psi$.  Then with probability $(p_{f'})^M$, they obtain $M$ copies of $\psi_f'$.
In the limit of large $M$, the parties $A$ and $B$ can, through LOCC, distill these copies
to $MS_f(A)>MNS_{\phi}(A)$ EPR pairs.

Thus $A$ and $B$ would be distilling more than $MNS_{\phi}(A)$ EPR pairs from $MN$ copies of $\phi$, and from
\cite{LP} their success probability must be $\exp(O(-MN))$, hence $p_f'\sim \exp(O(-N))$.
But for $E_{rnd}(\psi)>E_{sp}(\psi)$ under these circumstances
would require $S_f(A_XT^f_{XY})\sim \exp(O(N))$, which would require a forbidden increase in Schmidt number
across a bipartite split between group $A_XT^f_{XY}$ and all other parties.

Hence in the limit of large $N$, we cannot have advantageous random distillation of $N$ copies of a pure state. $\Box$.

\section{Conclusion}
We have generalised several of the results noted for specific cases in \cite{FL}.  We have more carefully defined what constitutes
``random'' distillation so that any apparent advantage in terms of entanglement yield is specifically due to the final
parties not being predetermined.   The advantageous random distillation
we previously noted for the $W$ and similar states has been shown to apply to the general $W$-class of three-qubit states
(and the GHZ class not to have a consistent property in this respect).  We have shown that for the important class of symmetric Dicke states
our $W$ protocol can
achieve conversions between states which are not achievable for predetermined
final parties.  Finally we have shown that advantageous random distillation does not occur in the many-copy limit, and hence is
a property specific to individual quantum states that cannot be considered in a regularised form, in contrast
to many other entanglement properties.

Clearly we have still only dealt with a limited class of states and the extremal conditions of a single copy or the many-copy limit.
Our quantitative approach does not readily generalise to all states - e.g. for random distillation to final states shared between more than two parties,
the lack of a standard entanglement measure makes the choice of target state more arbitrary, and an ``advantageous'' random distillation is less defined by a measure
than by the probability of achieving a given target state.  However, as demonstrated with Dicke states above, two-party entanglement measures can be used
to determine whether or not such states are achievable between predetermined parties.

For distillation to two-party entanglement from multiple copies of a state, an open question is how any advantage due to random distillation
scales with the number of copies, since we now know such advantage vanishes in the many-copy limit.

As noted in \cite{FL}, even when the target states are two-party and thus the final entanglement is reasonably well-defined,
the full ``structure'' of the output of random distillation would be defined by a probability distribution over final
entanglements for given pairs of parties, rather than a single number.  For example, the $W$ protocol for a $W$ state
shared between parties $A,B,C$ reliably produces an EPR pair between one of the three pairs of parties $AB, BC, AC$, with each
pairs having a probability of $1/3$ of receiving the EPR.  As shown in \cite{FL}, EPRs can be reliably produced from some $W$-like states which are not
symmetric, but in this case the probability of getting an EPR is not the same for each pair.  An interesting open question is what the optimum
such probability distribution (in terms of $E_{rnd}$) is for a given state, and how this can be determined from the form of the state. 

The authors acknowledge financial support from NSERC, CIFAR, the CRC program, CFI, OIT,
PREA, MITACS, CIPI and QuantumWorks.
\section{Appendix A}
Proof of Lemma 2 can be done algebraically as follows
\begin{widetext}
\begin{align}
q_{rnd}^2 - q(\rho_B)^2=&4\left(1-2\frac{\alpha^2}{\gamma^2}+\frac{\alpha^4}{\gamma^4}\right)\beta^2\gamma^2+4\alpha^4+\frac{\alpha^4\beta^4}{\gamma^4}
+8\alpha^2\beta\gamma\left(1-\frac{\alpha^2}{\gamma^2}\right)+4\frac{\alpha^4\beta^2}{\gamma^2}\nonumber\\
&+4\frac{\alpha^2\beta^3}{\gamma}\left(1-\frac{\alpha^2}{\gamma^2}\right)-4\beta^2(\alpha^2+\gamma^2)\\
=&\alpha^2\left[-12\beta^2+8\frac{\alpha^2\beta^2}{\gamma^2}+4\alpha^2+\frac{\alpha^2\beta^4}{\gamma^4}
+8\beta\gamma\left(1-\frac{\alpha^2}{\gamma^2}\right)
+4\frac{\beta^3}{\gamma}\left(1-\frac{\alpha^2}{\gamma^2}\right)\right]\\
=&\alpha^2\left[\beta^2\left(8\frac{\gamma}{\beta}+4\frac{\beta}{\gamma}-12\right)
+\alpha^2\left(8\frac{\beta^2}{\gamma^2}
+4+\frac{\beta^4}{\gamma^4}-8\frac{\beta}{\gamma}-4\frac{\beta^3}{\gamma^3}\right)\right]\\
=&\alpha^2\left[4\beta^2\left(\frac{\gamma}{\beta}-1\right)\left(2-\frac{\beta}{\gamma}\right)
+\alpha^2\left(\left(\frac{\beta^2}{\gamma^2}-\frac{2\beta}{\gamma}\right)^2
+4\left(1-\frac{\beta}{\gamma}\right)^2\right)\right]
\end{align}
\end{widetext}
There are thus 3 terms in the above.  We recall that $0<\alpha\le\beta\le\gamma$.
The first term is clearly $\ge 0$ since $\gamma\ge \beta$, and the other two terms are clearly $\ge 0$ since they are squared.
The first and third terms are both equal to 0 iff $\beta=\gamma$, but
in that case the second term is $>0$.  Thus
\begin{equation}
q_{rnd}>q(\rho_B)\quad\Box.
\end{equation}

\end{document}